\documentclass[aps,prd,preprint,nofootinbib]{revtex4}
\usepackage[dvips]{epsfig}
\usepackage{amsfonts,amssymb}

\newcommand{\be}{\begin{equation}}
\newcommand{\ee}{\end{equation}}
\newcommand{\ben}{\begin{displaymath}}
\newcommand{\een}{\end{displaymath}}
\newcommand{\bea}{\begin{eqnarray}}
\newcommand{\eea}{\end{eqnarray}}
\newcommand{\bean}{\begin{eqnarray*}}
\newcommand{\eean}{\end{eqnarray*}}

\def\l {\lambda}

\def\e {\epsilon}

\newcommand{\ads}[1]{\mbox{${AdS}_{#1}$}}

\newcommand{\tr}{\mbox{Tr}}

\newcommand{\commentout}[1]{}

\newcommand{\beq}{\begin{equation}}
\newcommand{\eeq}{\end{equation}}
\newcommand{\beqr}{\begin{displaymath}}
\newcommand{\eeqr}{\end{displaymath}}
\newcommand{\beqa}{\begin{eqnarray}}
\newcommand{\eeqa}{\end{eqnarray}}
\newcommand{\beqar}{\begin{eqnarray*}}
\newcommand{\eeqar}{\end{eqnarray*}}
\newcommand{\cN}{{\cal N}}

\newcommand{\cL}{{\cal L}}

\newcommand{\ti}[1]{\tilde{#1}}

\newcommand{\half}{\ensuremath{\frac{1}{2}}}

\newcommand{\N}[1]{\ensuremath{\cN=#1}}

\begin{document}

\title{\Large New open string solutions in $AdS_5$ }

\author{R. Ishizeki}
\email{rishizek@purdue.edu}
\author{M. Kruczenski}
\email{markru@purdue.edu}
\author{A. Tirziu}
\email{atirziu@purdue.edu}
\affiliation{ Department of Physics, Purdue University,
525 Northwestern Ave., W. Lafayette, IN 47907-2036, USA.
}
\date{\today}


\begin{abstract}
We describe new solutions for open string moving in \ads{5} and ending in the boundary, namely dual to Wilson loops in \N{4} SYM theory.
First we introduce an ansatz for Euclidean curves whose shape contains an arbitrary function. They are BPS and the dual surfaces can be found
exactly. After an inversion they become closed Wilson loops whose expectation value is $W=\exp(-\sqrt{\lambda})$. After that we consider several
Wilson loops for \N{4} SYM in a pp-wave metric and find the dual surfaces in an \ads{5} pp-wave background. Using the fact that the pp-wave is
conformally flat, we apply a conformal transformation to obtain novel surfaces describing strings moving in AdS space in Poincare coordinates
and dual to Wilson loops for \N{4} SYM in flat space.
\end{abstract}


\maketitle

\section{Introduction}
\label{intro}

The understanding of the AdS/CFT correspondence between $\mathcal{N}=4$ SYM theory and the dual type IIB string theory on $AdS_5 \times S^{5}$
has seen steady progress in recent years.
Recently, the study of the correspondence between open string solutions and the dual Wilson loops received renewed attention. Following earlier
discussions \cite{k,m} on the cusp anomaly, in \cite{am1} the authors proposed that the planar four gluon scattering amplitude is described,
at strong coupling, by an open string solution ending on a sequence of four light-like lines at the boundary of $AdS_5$ space in $T$-dual coordinates.
This strong coupling string calculation is in agreement with the gauge theory conjecture of \cite{bds}. A series of papers studied various aspects of this
proposal \cite{krtt,b,kr,ms}\footnote{See also \cite{alday} and the references therein.}. The proposal put forward in \cite{am1} was further analyzed in \cite{am2}. There, a prescription of how to compute
amplitudes for processes that involve local operators and gluon final states was also proposed. While the string ends on light-like lines at the
horizon in ordinary $AdS_5$ space, the operator insertion leads to string solutions going to the boundary of ordinary $AdS_5$ space, and which,
at the same time are stretched in the direction in which the momentum of the operator is non-trivial.

Motivated by these recent developments in understanding various gauge theory scattering processes at strong coupling through a string description,
we find in this paper several new open string solutions in $AdS_5$ that end at the boundary on various Wilson lines.

First we find a class of solutions ending on space-like Wilson loops. These solutions extend to the horizon and end at the boundary
of $AdS_5$ on a large class of open lines. The (regularized) classical area of the corresponding world-sheet is zero since these solutions are BPS.
 Performing an inversion transformation we obtain a class of solutions that end on a closed space-like Wilson loop at the boundary.
In this case their expectation values turn out to be  $\langle W\rangle =e^{-\sqrt{\lambda}}$.  This result is in agreement with the conjecture
in \cite{dg} that considers two Euclidean Wilson $W_{1,2}$ loops which are related by an inversion, with one of them $W_1$ extending to infinity
and the other $W_2$ given by a closed contour.  In that case the conjecture is that $\langle W_2 \rangle = e^{-\sqrt{\lambda}} \langle W_1 \rangle$ at the leading order in strong coupling $\lambda$.
The usual example is the straight-line $\langle W_1\rangle =1$ and the circle $\langle W_2 \rangle =e^{-\sqrt{\lambda}}$. It would we interesting to obtain the relationship between the Wilson loops $W_{1,2}$ also at subleading orders in strong coupling.

 In the second part of this paper we exploit the relation between the $AdS_5$ $pp$-wave and pure $AdS_5$ to find new open string solutions in the latter space.
We extend the solution obtained in \cite{KT} in this context to other, more complex solutions. We find solutions ending on two straight light-like lines
in the $x_{+}$ direction on the boundary in the $pp$-wave $AdS_5$ space, which when translated in the pure $AdS_5$ look like hyperbolas in a ($t,x$) plane.
Another new solution that we find is one that ends on a time-like straight line in the time direction. In pure $AdS_5$ coordinates it looks like a tangent
shaped line in a ($x_{\pm}=\frac{1}{\sqrt{2}}(x\pm t)$) plane. The function $z=z(x)$ for fixed $t$ is a line with one end on the boundary, and the other
at the horizon. We suggest that this type of solution can describe the scattering between a quark at the boundary moving at a speed less than the speed
of light and a gluon coming out from infinity. However, it corresponds to forward scattering. It should be desirable to look for other solutions in
which the incoming gluon changes direction. We find also solutions that end on two parallel lines in the time direction in the $pp$-wave, which are
mapped into two tangent type Wilson lines on the boundary in pure $AdS_5$ space.

 The organization of this paper is as follows: In the next section we described the new Wilson loops solutions found using an ansatz in light-cone
coordinates. In section \ref{pp-wave} we describe the solutions found by conformally mapping the pp-wave onto flat space. Finally we
give our conclusions in the last section.

\section{String solutions ending on open/closed space-like Wilson loops. }\label{xploops}

The Nambu-Goto action for a string moving in $AdS_5$ in Poincar\'e coordinates is given by
\beq
 S = i\frac{\sqrt{\l}}{2\pi} \int d\sigma d\tau \cL
   = i\frac{\sqrt{\l}}{2\pi} \int d\sigma d\tau \frac{1}{z^2}
                             \sqrt{ (\partial_\sigma X)^2 (\partial_\tau X)^2
                                    -(\partial_\sigma X.\partial_\tau X)^2}
\label{SstNG}
\eeq
The space-time metric is set as (here $x_{\pm}=\frac{1}{\sqrt{2}}(x\pm t)$)
\beq
 ds^2 = \frac{1}{z^2}\left( 2 dx_+ dx_- + dx_1^2 + dx_2^2 + dz^2 \right)
\label{metric_xp}
\eeq
We consider the following ansatz
\begin{equation}
x_+ = u(\sigma,\tau) ,  \quad
x_- = 0 , \quad
x_1 = \tau  , \quad
x_2 = a  , \quad
z = \sigma \label{gfs}
\end{equation}
where $a$ is constant which we can set to zero (but we leave it here for later use). We also have that
$z=\sigma \geq 0$, while $\tau$ varies over the real line. The surface ends at the boundary $z=\sigma=0$ in the (space-like) line
\beq
x_+ = u(\sigma=0,\tau) ,  \quad x_- = 0 , \quad x_1 = \tau  , \quad x_2 = a.
\eeq
 Our objective is to find $u(\sigma,\tau)$ given an arbitrary function $u(\sigma=0,\tau)$.

\commentout{
Then, we can obtain the following equations of motion for (\ref{SstNG}),
\beqa
 \partial_\sigma \frac{\partial \cL}{\partial x'_+} + \partial_\tau \frac{\partial \cL}{\partial \dot{x}_+}
 &=& \frac{\partial \cL}{\partial x_+} \label{equ}  \\
 \partial_\sigma \frac{\partial \cL}{\partial x'_-} + \partial_\tau \frac{\partial \cL}{\partial \dot{x}_-}
 &=& \frac{\partial \cL}{\partial x_-} \label{eqv}  \\
 \partial_\sigma \frac{\partial \cL}{\partial w'} + \partial_\tau \frac{\partial \cL}{\partial \dot{w}}
 &=& \frac{\partial \cL}{\partial w} \label{eqw}  \\
 \partial_\sigma \frac{\partial \cL}{\partial y'} + \partial_\tau \frac{\partial \cL}{\partial \dot{y}}
 &=& \frac{\partial \cL}{\partial y} \label{eqy}  \\
 \partial_\sigma \frac{\partial \cL}{\partial z'} + \partial_\tau \frac{\partial \cL}{\partial \dot{z}}
 &=& \frac{\partial \cL}{\partial z} \label{eqz}
\eeqa }
Solving the equation of motion for $x_-$, we obtain,
\beq
  u''(\sigma,\tau) - \frac{2}{\sigma}u'(\sigma,\tau) + \ddot{u}(\sigma,\tau) = 0
\label{v_sol}
\eeq
 We also find that the equations for all the other coordinates are satisfied as long as eq.(\ref{v_sol}) is true.
A simple check is that $u=0$ is a solution corresponding to a straight line along $x_2$.
The main point of this ansatz is that the resulting equation (\ref{v_sol}) is linear and therefore can be solved
using a variety of standard methods for any initial condition $u(\sigma=0,\tau)$.
Indeed, taking the Fourier transform (where we use the fact that $u(\sigma,\tau)$ is real):
\begin{equation}
  u(\sigma,\tau) = \int_0^\infty d\omega \left[e^{i\omega \tau} u(\sigma,\omega)+e^{-i\omega \tau} u^*(\sigma,\omega)\right]
\end{equation}
in (\ref{v_sol}) leads to
\begin{equation}
  u''(\sigma,\omega) - \frac{2}{\sigma}u'(\sigma,\omega) - \omega^2 u(\sigma,\omega) = 0
\label{v_sol2}
\end{equation}
The general solution of this equation is
\begin{equation}
  u(\sigma,\omega) = C_1 e^{\omega\sigma} (\omega\sigma-1)
             + C_2 e^{-\omega\sigma} (\omega\sigma+1)
\end{equation}
Since $\omega>0$ we keep the second solution which does not diverge as $\sigma\rightarrow\infty$.
After that, the most general solution of (\ref{v_sol}) is
\beq
 u(\sigma,\tau)
  =  \int_0^\infty d\omega  e^{-\omega\sigma} (\omega\sigma+1) \left[ u(0,\omega) e^{i\omega\tau}+ u^*(0,\omega) e^{-i\omega\tau}\right]
\label{u_sol}
\eeq
which is the main result of this section, allowing to compute $u(\sigma,\tau)$ given $u(0,\tau)$.
 A simpler expression can be obtained if we define the positive frequency part of $u(\sigma,\tau)$ as
\beq
 u(\sigma,\tau) = \chi_+(\sigma,\tau) + \chi_+^*(\sigma,\tau), \ \ \ \ \ \chi_+(\sigma,\tau)=  \int_0^\infty d\omega e^{i\omega \tau} u(\sigma,\omega)
\label{uchi}
\eeq
 Now we should find $\chi_+(\sigma,\tau)$ given $\chi_+(0,\tau)$. Using the previous calculations we find
\beqa
 \chi_+(\sigma,\tau) &=&   \int_0^\infty d\omega  e^{i\omega\tau-\omega\sigma} (\omega\sigma+1) u(0,\omega) \\
                     &=&   \left. (1-\partial_\xi) \int_0^\infty d\omega  e^{i\omega\tau-\xi\omega\sigma}  u(0,\omega) \right|_{\xi=1} \\
                     &=&   \chi_+(\tau+i\sigma) - i\sigma \chi_+'(\tau+i\sigma)
\eeqa
where $\chi_+'(\tau)=\partial_\tau \chi_+(\tau)$.
 In fact it is straight-forward to check that the function
\beq
  \chi_+(\sigma,\tau) = \chi_+(\tau+i\sigma) - i\sigma \chi_+'(\tau+i\sigma)
\label{chisol}
\eeq
satisfies eq.(\ref{v_sol}). As an example consider a bell-shaped Wilson loop
\beq
 x_+ = \frac{1}{\tau^2+a^2}, \ \ \ x_2=\tau, \ \ \ x_-=0
\eeq
 Doing the Fourier transform we obtain (assuming $a>0$):
\beq
 \frac{1}{\tau^2+a^2} = \frac{1}{2a} \int_0^\infty e^{-a\omega+i\omega\tau} d\omega + \mbox{c.c.} ,
 \ \ \ \Rightarrow \ \ \ \chi_+(0,\tau) = \frac{1}{2a}\frac{1}{a-i\tau}
\eeq
which implies
\beq
 \chi_+(\sigma,\tau) = \chi_+(\tau+i\sigma) - i\sigma \chi_+'(\tau+i\sigma) = \frac{1}{2a}\frac{a+2\sigma-i\tau}{(a-i\tau+\sigma)^2}
\eeq
From here, using eq.(\ref{uchi}), we find the solution
\beq
 u(\sigma,\tau) = \frac{(a+\sigma)^2(a+2\sigma)+a\tau^2}{a\left[(a+\sigma)^2+\tau^2\right]^2}
\eeq
 This is just an example to check the validity of the method. Given any shape we can write the solution in term of Fourier integrals
using eq.(\ref{u_sol}) or (\ref{chisol}).

 Going back to the general solution (\ref{gfs}), the action $S$ is imaginary and proportional to the area of the worldsheet $A$:
\beq
S= i A \frac{\sqrt{\lambda}}{2 \pi}
\eeq
The action is
\begin{equation}
 -i S = \frac{\sqrt{\l}}{2\pi} \int d\sigma d\tau \cL
   = \frac{\sqrt{\l}}{2\pi} \int \frac{d\sigma d\tau}{\sigma^2} = \frac{\sqrt{\lambda}}{2 \pi} \frac{T}{\epsilon}
\label{SstAreaqa}
\end{equation}
where we chose the $\tau$ interval $-\frac{T}{2}\leq \tau \leq\frac{T}{2}$ and $T$ is large. Also, we introduced a cutoff, $\epsilon$,
in $z$ to account for the divergency near the boundary. The solution (\ref{gfs}) at the boundary $z=0$ is an open line as $\tau$ spans
the real line. While the action is divergent, one should recall \cite{dgo} that in fact the physical result is obtained by adding a boundary
term which precisely cancels the linear divergency $\frac{1}{\epsilon}$, so the physical action is actually zero. This suggests that this solution
is BPS as we now check. Note also that the result is independent of the function $u(\tau, \sigma)$.

\subsection{Supersymmetry of the Wilson loop}

 The Wilson loop we considered in the previous subsection is BPS as we now proceed to prove. The arguments are well known, here we follow
\cite{Zarembo}.
Given the Wilson loop
\beq
W = \frac{1}{N} \tr \hat{P} \exp\int d\tau \left(iA^\mu(x)\dot{x}_\mu + \Phi_1(x) |\dot{x}|\right)
\eeq
where $x_\mu(\tau)$ parameterizes the path and $\dot{x}=\partial_\tau x$. Also, $\Phi_1$ is a scalar field of the \N{4} SYM theory.
 Supersymmetry is preserved if we can find spinor solutions $\epsilon$ to satisfy
\beq
 \left(i\Gamma^\mu \dot{x}_\mu + \tilde{\Gamma}_1 |\dot{x}|\right)\epsilon =0
\eeq
 where $\Gamma^{\mu=\pm,1,2}$, $\tilde{\Gamma}^{i=1 \cdots 6}$ are ten-dimensional gamma matrices. In the case of the path analyzed in the previous
subsection, this boils down to
\beq
 \left(i\Gamma^+ \dot{x}_+ + i \Gamma^1 + \tilde{\Gamma}^1\right)\epsilon=0
\eeq
 Since we are taking $x_+(\tau)$ as an arbitrary function we require (in special cases such as $x_+=0$ or $x_+=\tau$ the Wilson loop is $\half$ BPS):
\beq
 \Gamma^+ \epsilon =0, \ \ \ \ \left(i \Gamma^1 + \tilde{\Gamma}^1\right) \epsilon =0
\eeq
 Using for example the representation of gamma matrices in terms of creation an annihilation operators (see \cite{Polchinskibook}) it is
easily seen that only four independent solutions exist, namely the Wilson loop preserve one quarter of the sixteen supercharges. In other
words these Wilson loops are $\frac{1}{4}$ BPS. Their expectation value should be $W=1$ independent of the coupling.
 Notice that in \cite{Zarembo} it was stated that Wilson loops involving just one of the scalar fields (here $\Phi_1$) can be BPS only
if they are straight lines. However the analysis was done in Euclidean space, in the Minkowski case this is no longer true as we have just shown.

\subsection{Inversion and closed Wilson loop}

To construct a closed loop at the boundary of $AdS_5$ we consider the inversion transformation $\tilde{x}_{\mu}=\frac{x_{\mu}}{|x|^2}$
which leaves the metric (\ref{metric_xp}) invariant. After this transformation the solution (\ref{gfs}) becomes
\begin{equation}
 \ti{x}_+ = \frac{u(\sigma,\tau)}{\tau^2+\sigma^2+a^2}\nonumber
\end{equation}
\begin{equation}
 \quad \ti{x}_- = 0 \ ,  \quad \ti{x}_1 = \frac{\tau}{\tau^2+\sigma^2+a^2} \ ,
\quad \ti{x}_2 = \frac{a}{\tau^2+\sigma^2+a^2} \ ,
\quad \ti{z} = \frac{\sigma}{\tau^2+\sigma^2+a^2} \label{mas}
\end{equation}
Starting with the Nambu-Goto action in the new coordinates with metric
\beq
 ds^2 = \frac{1}{\tilde{z}^2}\left(2 d \tilde{x}_+ d \tilde{x}_- + d \tilde{x}_1^2 + d \tilde{x}_2^2 + d \tilde{z}^2 \right)
\label{metric_xpt}
\eeq
one can easily see that indeed (\ref{mas}) is a solution with $u$ satisfying again (\ref{v_sol}).
Before doing the inversion transformation, the radial coordinate $z$ goes all the way to the horizon $0\leq z <\infty$.
After the inversion $\tilde{z}$ has a maximum attained at $\tau=0$, and $\sigma=a$, so $0\leq \tilde{z}\leq \frac{1}{2 a}$.
Thus after the inversion the surfaces closes up in the bulk (here we assume $a\neq 0$). More precisely,
the solution (\ref{mas}) projected over the subspace $x_+=x_-=0$ defines a sphere
\begin{equation}
(\tilde{x}_2-\frac{1}{2a})^2+\tilde{x}_1^2+\tilde{z}^2=\frac{1}{4 a^2}
\end{equation}
which is the same as the circular solution in Euclidean space considered in \cite{bcfm,dg,Erickson:2000af}. Over that sphere $x_+$ varies according
to the function $x_+=u(\sigma,\tau)$.

At the boundary, $\tilde{z}=0$, i.e. $\sigma=0$, we obtain
\begin{equation}
 \ti{x}_+ = \frac{u(\sigma=0,\tau)}{\tau^2+a^2}, \quad
 \ti{x}_- = 0, \quad
 \ti{x}_1 = \frac{\tau}{\tau^2+a^2}  \quad
 \ti{x}_2 = \frac{a}{\tau^2+a^2},
\end{equation}
Projected onto the subspace $x_-,x_+=0$ this curve is a circle with radius, $\frac{1}{2a}$,
\begin{equation}
 \ti{x}_1^2 + \left( \ti{x}_2-\frac{1}{2 a} \right)^2 = \frac{1}{4a^2}
\end{equation}
Therefore, we can parametrize the solution at the boundary as follows
\begin{eqnarray}
  \ti{x}_+ = \frac{\left(1+\cos \theta \right) u(\theta)}{2a^2}, \quad
  \ti{x}_1 = \frac{1}{2a} \sin \theta, \quad
  \ti{x}_2 = \frac{1}{2a} + \frac{1}{2a} \cos \theta,
\end{eqnarray}
where $u(\theta) = u(\sigma=0,\tau)$, $\tau=\tau(\theta)= a \tan \frac{\theta}{2}$, and $0\leq \theta< 2\pi$.
As a result of the inversion transformation the Wilson lines closes up. At least as long as we take $x_+=u(0,\tau)$ as a
bounded function.

Let us compute the action for the solution (\ref{mas})
\begin{equation}
 - i S = \frac{\sqrt{\l}}{2\pi} \int d\sigma d\tau \cL
   = \frac{\sqrt{\l}}{2\pi} \int \frac{d\sigma d\tau}{\sigma^2}
\label{SstArea}
\end{equation}
As in the original coordinates, in the inverted coordinates we again introduce a cutoff near the boundary of $AdS_5$.
This cutoff is now in $\tilde{z}$.  This leads to the condition
\begin{equation}
 \ti{z} = \frac{\sigma}{\tau^2+\sigma^2+a^2} > \epsilon
\end{equation}
This equation is equivalent to
\begin{equation}
   \tau^2 + \left( \sigma - \frac{1}{2\e} \right)^2 < \frac{1}{4\e^2}-a^2
\end{equation}
Now we can parametrize such region with
\begin{eqnarray}
  \tau &=& \rho \sin \phi   \\
  \sigma &=& \rho \cos \phi  + \frac{1}{2\e}
\end{eqnarray}
where $0 \le \rho < R=\sqrt{\frac{1}{4 \epsilon^2} - a^2}$, and
$0\le \phi < 2\pi$.

 Therefore the action (\ref{SstArea}) can be computed using this parametrization as
\begin{eqnarray}
 -i S &=& \frac{\sqrt{\l}}{2\pi} \int \frac{\rho d\rho d\phi}
          {\left(\rho \cos \phi +\frac{1}{2\e} \right)^2}  \\
   &=& \frac{\sqrt{\l}}{2a\e} - \sqrt{\l} + O(\e^2)  \label{lnm}
\end{eqnarray}
As in the case of the open Wilson line, the linear divergent part is proportional to the length of the string in the ($\tilde{y},\tilde{w}$)
plane at the boundary. When the lengths in the two cases are equal, i.e. $T=\frac{\pi}{a}$, it makes sense to subtract from (\ref{lnm}) the
open line result so that the linear divergent term cancels.  In contrast to the action (or worldsheet area) of the open loop in (\ref{SstAreaqa}),
the action for the closed loop in (\ref{lnm}) has a non-trivial finite part. This relationship is the same as the one from open/closed Wilson loops
analyzed in Euclidean space in \cite{dg}. One can also compute string $1$-loop corrections to these solutions following the procedure developed
in \cite{kt}. Let us again observe that the result in (\ref{lnm}) is independent of the function $u(\tau,\sigma)$. This Wilson loop is not BPS
but should be invariant under a superconformal charge. Presumably the expectation value of this class of Wilson loops can be computed exactly by using the methods of \cite{Pestun}
(generalized to Minkowski signature). Summing rainbow diagrams also leads to the same result as in the case of the circular Wilson loop. Finally, let us also mention that it would be interesting to see whether there exists any relationship between the class of solutions considered here and the ones discussed in \cite{Mikhailov}.

\subsection{General procedure}

 In this subsection we show that the same procedure can in principle be applied to other solutions with Euclidean world-sheet metric.
Indeed, suppose we have a metric of the form
\beq
 ds^2 = 2 g_{+-}(x_i,x_\pm) dx_+ dx_- + g_{ij}(x_l) dx^i dx^j
\eeq
and a (conformal gauge) solution $x_i=\bar{x}_i(\sigma,\tau)$ with $x_\pm =0$.
Now let us look for a new solution with the ansatz $x_+=u(\sigma,\tau)$, $x_-=0$, and $x_i=\bar{x}_i(\sigma,\tau)$.
The constraints (recalling that we are looking at Euclidean solutions)
\beq
 g_{\mu\nu} \left(\dot{x}^\mu \dot{x}^\nu - x'^\mu x'^\nu \right) = 0 , \ \   g_{\mu\nu} \dot{x}^\mu x'^\nu =0
\eeq
are automatically satisfied (if they were satisfied for the original solution). The equations of motion
\beq
  \partial_\tau   \left( g_{\mu\nu} \dot{x}^\nu \right)
+ \partial_\sigma \left( g_{\mu\nu} x'^\nu      \right) =
  \partial_\mu g_{\alpha\beta} \left( \dot{x}^\alpha \dot{x}^\beta + x'^\alpha x'^\beta\right)
\eeq
 are satisfied when $\mu=i$, and, since $\partial_{\pm}g_{ij}=0$, also for $\mu=+$. The equation with $\mu=-$ gives
\beq
 \partial_\tau \left(g_{-+} \dot{x}^+\right)
+ \partial_\sigma \left(g_{-+} x'^+\right) = 0
\eeq
After replacing for the solution $x_i=\bar{x}_i(\sigma,\tau)$, this equation of motion is linear in $x_{+}$ if $\partial_+ g_{+-}=0$ as was the case in our previous discussion. If not, the equation is in general non-linear.
In both cases this method can provide new solutions given old Euclidean ones. Here we looked at the straight string and the circular
Wilson loop. It might be interesting to analyze other cases as well.

\section{Open string solutions obtained through the map to $pp$-wave background}\label{pp-wave}

One can find new open string solutions in $AdS_5$ by exploiting the map between pure $AdS_5$ and the $AdS_5$ $pp$-wave backgrounds.
These solutions correspond to Wilson loops in Minkowski space in the dual gauge theory. The relationship between the $AdS_5$ metric
(here $x_{\pm}= \frac{x \pm t}{\sqrt{2}}$, $i=1,2$, and we use the notation $x_{+}\equiv x^{+}$, $x_{-}\equiv x^{-}$)
\begin{equation}
ds^2 = \frac{1}{\tilde{z}^2}(2 d \tilde{x}_{+} d \tilde{x}_{-}+ d \tilde{x}_i d\tilde{x}_i + d\tilde{z} d \tilde{z})
\end{equation}
and $pp$-wave $AdS_5$ metric
\begin{equation}
ds^2= \frac{1}{z^2} ( 2 d x_{+} d x_{-} - \frac{1}{2}(x_i^2 + z^2) F(x_{+})d x_{+}^2 + d x_i^2 + dz^2)
\end{equation}
is given by \cite{bcr}
\begin{equation}
\tilde{x}_{+}= f(x_{+}),  \quad \tilde{x}_{-}= x_{-} - \frac{1}{4}(x_i^2+ z^2) \frac{f''}{f'}, \quad \tilde{x}_i = x_i \sqrt{f'}, \quad \tilde{z}= z \sqrt{f'}
\end{equation}
where $F$ is given by the Schwarzian derivative
\begin{equation}
F(x_{+})=\{f,x_{+}\}= \frac{f'''}{f'}- \frac{3}{2}\bigg(\frac{f''}{f'}\bigg)^{2} \label{wjm}
\end{equation}
Mapping back simple solutions in the $pp$-wave $AdS_5$ background usually produce new and potentially useful solutions in pure $AdS_5$.

 Although in principle one can consider various functions $F(x_+)$ to obtain new Wilson loops, in this paper we just consider the simplest
non-trivial case corresponding to $F(x_+)=a$ where $a$ is a constant. In that case, the solution to (\ref{wjm}) has
the form (up to shifts\footnote{Note that a shift in $x_{+}$ is an isometry if $F(x_{+})$ is constant.} in $x_{+}$)
\begin{equation}
f(x_{+})= A \sqrt{\frac{2}{a}}\tan \sqrt{\frac{a}{2}}x_{+}
\end{equation}
where $A$ is a constant. Notice that in the limit $a\rightarrow 0$, $AdS_5$ is mapped onto itself through
\begin{equation}
f(x_{+})= \frac{A}{x_{+}}
\end{equation}
 The transformation of coordinates in this case is given by
\begin{equation}
\tilde{x}_{+}=\frac{A}{x_{+}}, \quad \tilde{x}_{-}=x_{-}+ \frac{1}{2}\frac{x_i^2+ z^2}{x_{+}},
\quad \tilde{x}_i= x_i \frac{\sqrt{-A}}{x_{+}}, \quad \tilde{z}=z \frac{\sqrt{-A}}{x_{+}}
\end{equation}
where one needs to choose $A<0$.

 In what follows we will be concerned with the situation when $F({x_{+}})$ is constant but non-zero. As in \cite{KT} let us choose
$f(x_{+})$ of the form $f(x_{+})=\frac{1}{\mu}\tan \mu x_{+}$. Therefore, the $AdS_5$ $pp$-wave metric and the transformation are
\begin{equation}
ds^2= \frac{1}{z^2} ( 2 d x_{+} d x_{-} - \mu^2 (x_i^2 + z^2) d x_{+}^2 + d x_i^2 + dz^2)  \label{owd}
\end{equation}
and
\begin{equation}
\tilde{x}_{+}= \frac{1}{\mu}\tan \mu x_{+},  \quad \tilde{x}_{-}= x_{-} - \frac{1}{2}\mu(x_i^2+ z^2) \tan \mu x_{+}, \quad \tilde{x}_i = \frac{x_i}{\cos \mu x_{+}}, \quad \tilde{z}= \frac{z}{\cos \mu x_{+}}  \label{tra}
\end{equation}
The parameter $\mu$ can be scaled away by a rescaling $x_{+}\rightarrow \frac{x_{+}}{\mu}$, $x_{-}\rightarrow \mu x_{-}$, so it can be set to $1$. However, below we will keep $\mu$ explicitly in order to be able to make the connection to pure $AdS_5$ which corresponds to the limit $\mu \rightarrow 0$. Using the map (\ref{tra}) we find below new open string solutions.
Let us finish this section by pointing out that the map (\ref{tra}) is singular at $\mu x_{+}=(2k+1) \frac{\pi}{2}$, $k\in \mathbb{Z}$. This means that this map takes part of the $pp$-wave $AdS_5$ space into the full $AdS_5$ space.

\subsection{Straight lines along the $x_{+}$ direction}

\subsubsection{$1$-line}

We start by reviewing the solution in the $AdS_5$ $pp$-wave  that ends on a single light-like solution, and was found in \cite{KT}
\begin{equation}
x_{+}=\tau, \quad x_{-}=0, \quad x_i=0, \quad z=\sigma  \label{kng}
\end{equation}
On the boundary $\sigma=0$, this solution may be interpreted as the world-line of a massless quark at $x_{-}=0$. Let us point out that this is also a solution in pure $AdS_5$ whose worldsheet is just a half-plane ending on the $x_{+}$ line at the boundary.
One may wonder whether there exists a direct extension of this solution to a time-like line with $x_i \neq 0$ but constant, as for example $x_1=b$. It turns out that there are no such solutions, i.e. $b=0$. The charges in the $AdS_5$ $pp$-wave  background have been computed in \cite{KT}, where it was found that their divergent parts are related by
\begin{equation}
P_{+}\approx \frac{\lambda}{4 \pi}\ln |P_{-}|
\end{equation}

\begin{figure}
\centerline{\includegraphics[scale=1]{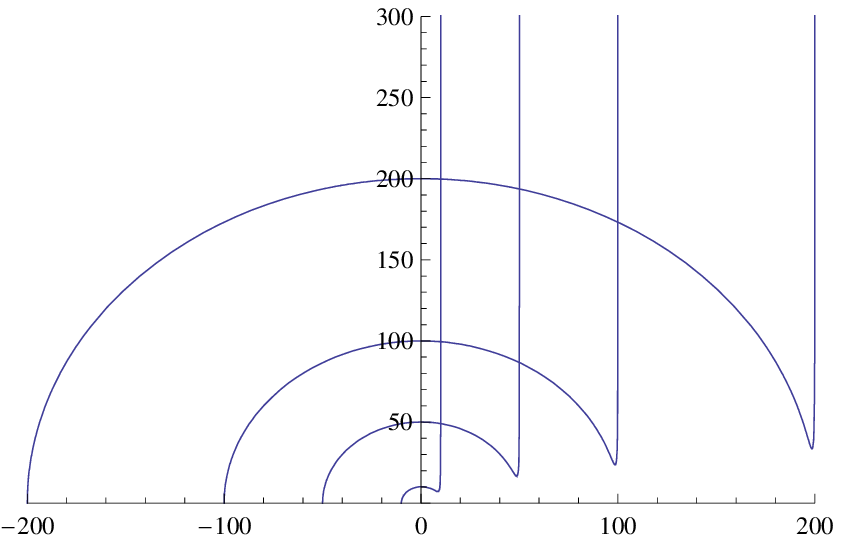}}
\caption{Plot of $\tilde{z}$ versus $\tilde{x}$ for $\mu=1$, and $\tilde{t}=- 200,-100,-50,-10$.}
\label{fig1}
\end{figure}

\begin{figure}
\centerline{\includegraphics[scale=1]{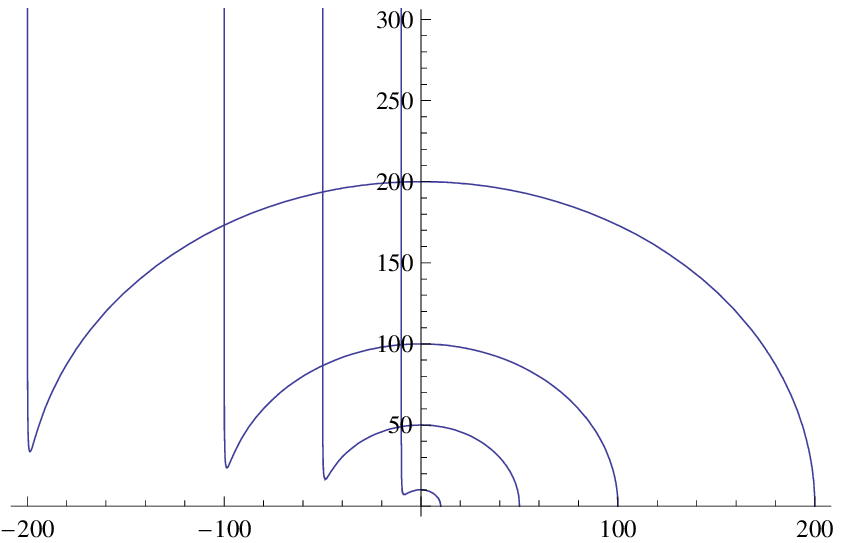}}
\caption{Plot of $\tilde{z}$ versus $\tilde{x}$ for $\mu=1$, and $\tilde{t}=200, 100,50,10$.}
\label{fig2}
\end{figure}

Let us use the transformation (\ref{tra}) and write the solution (\ref{kng}) in pure $AdS_5$
\begin{equation}
\tilde{x}_{+}=\frac{1}{\mu}\tan \mu \tau, \quad \tilde{x}_{-}=-\frac{\mu}{2}\sigma^2\tan \mu \tau, \quad \tilde{z}=\frac{\sigma}{\cos \mu \tau}, \quad \tilde{x}_i=0
\end{equation}
The worldsheet surface is given by
\begin{equation}
\tilde{z}=\sqrt{-\frac{2 \tilde{x}_{-}}{\mu^2 \tilde{x}_{+}}(1+ \mu^2 \tilde{x}_{+}^2)}  \label{lnd}
\end{equation}
On the boundary this solution again ends on a light-like line but now in contrast with (\ref{kng}) the string worldsheet is more complicated. For convenience let us parametrize the surface as
\begin{equation}
\tilde{z}(\tilde{x},\tilde{t})=\sqrt{-\frac{2}{\mu^2}\frac{\tilde{x}-\tilde{t}}{\tilde{x}+\tilde{t}}(1+ \frac{\mu^2}{2}(\tilde{x}+\tilde{t})^2)}
\end{equation}
The plot of $\tilde{z}$ as a function of $\tilde{x}$ for fixed $\tilde{t}<0$ is shown in Figure \ref{fig1}, while for $\tilde{t}>0$ in Figure \ref{fig2}.
For $\tilde{t}<0$ the rounded spike is to the left of the vertical straight line and it moves to the left as $\tilde{t}$ increases.  At $\tilde{t}=0$ the curve is a straight vertical line. As $\tilde{t}$ becomes positive the spike moves further to the left being now to the right of the vertical line.
The position of the maximum of the semi-circular part of the solution depends on $\tilde{t}$ and it is at zero in the $|\tilde{t}|\rightarrow\infty$ limit.

As already pointed out, the position of the rounded spike is close to $-\tilde{t}$ for $\tilde{t}<0$ but a bit to the left of the vertical straight line  (see figure \ref{fig1}). This shift depends on $\tilde{t}$ and in the large $\tilde{t}$ limit we obtain $\tilde{x}_s= - \tilde{t}- \delta_1$, with $\delta_1=\frac{\sqrt{2}}{\mu}$. For $\tilde{t}>0$ the position of the spike is a little bit to the right of the vertical line (figure \ref{fig2}). In the limit $\tilde{t}\rightarrow  \infty$, we now have
$\tilde{x}_s= - \tilde{t}+ \delta_1$. This means that there is a time delay in this process when moving from $-\infty$ to $\infty$ in $\tilde{t}$. In the limit $|\tilde{t}| \rightarrow \infty$ the height of the spike goes to the horizon as $z_s \sim 2^{\frac{5}{4}}\sqrt{\frac{|\tilde{t}|}{\mu}}$. The plots in figure \ref{fig1},\ref{fig2}, suggest that we can think of this solution as corresponding to a scattering process in the $1+1$ dimensional world-sheet with $\tilde{t},\tilde{x}$ being the $2d$ time and space coordinates. Because of the symmetry of the solution under $(\tilde{x},\tilde{t})\rightarrow (-\tilde{x},-\tilde{t})$ this is an elastic process, which is characteristic to integrable models in $1+1$ dimensions.

 Regarding the space-time/gauge theory interpretation of this solution, it was suggested in \cite{KT} that one can identify the end of the string
attached to the boundary as a quark. In addition there is the spike coming out from the horizon, which may be identified with a gluon. Then what
this solution appears to describe is a quark-gluon scattering process, but it is not clear how to  make this suggestion more precise. It would be
easier if one could find solutions describing scattering in which the gluon changes direction\footnote{In that respect, see \cite{spikes} for some
examples of how to construct solutions involving spikes.}.

 For completeness it is interesting to compute the energy and momentum of the solution. Since one is forcing the quark to move along a particular
trajectory, they are not conserved. In fact, the result determines the force necessary to push the quark. For $\tilde{t}>0$ the energy is
$E=\frac{\sqrt{\lambda}}{2 \pi}\mathcal{E}$
\begin{eqnarray}
\mathcal{E}(\tilde{t})&=& \int_{-\tilde{t}}^{\tilde{t}}d \tilde{x} \frac{\partial L}{\partial \dot{\tilde{t}}}=\int_{-\tilde{t}+\epsilon_0}^{\tilde{t}-\epsilon_0}d \tilde{x} \frac{1+ \tilde{z}^{'^2}}{\tilde{z}^2 \sqrt{1- \dot{\tilde{z}}^2+ \tilde{z}^{'^2}}}
=\frac{\mu}{4 \sqrt{2}}\bigg[-1+ \frac{4}{2+ \mu^2 \epsilon_0^2}+\frac{4}{2+ \mu^2 (\epsilon_0 - 2 \tilde{t})^2}\nonumber\\
&+&\frac{\epsilon_0}{\epsilon_0-2 \tilde{t}}+\frac{2 \tilde{t} }{\epsilon_0}+ 2 \log \frac{2 \tilde{t}-\epsilon_0}{\epsilon_0}\bigg]=\frac{\mu \tilde{t}}{2 \sqrt{2}\epsilon_0}+\frac{\mu}{4\sqrt{2}}\bigg[-3 + \frac{2}{1+ 2 \mu^2 \tilde{t}^2}- \log \frac{\epsilon_0^2}{4 \tilde{t}^2}\bigg] \label{jes}
\end{eqnarray}
where dot represents derivative in respect to $\tilde{t}$, and prime is derivative in respect to $\tilde{x}$. We have introduced a small cutoff $\epsilon_0$, and expanded in the last line in small $\epsilon_0$. The cutoff $\epsilon_0$ can be related to the physical cutoff $\epsilon$ in $\tilde{z}$ by
$\epsilon=\sqrt{2 \tilde{t}+\frac{1}{\mu^2 \tilde{t}}}\sqrt{\epsilon_0}$. For $\tilde{t}<0$ one obtains a similar expression
\begin{equation}
\mathcal{E}(\tilde{t})= -\frac{\mu \tilde{t}}{2 \sqrt{2}\epsilon_0}+\frac{\mu}{4\sqrt{2}}\bigg[-3 + \frac{2}{1+ 2\mu^2 \tilde{t}^2}- \log \frac{\epsilon_0^2}{4 \tilde{t}^2}\bigg]+O(\epsilon_0)
\end{equation}
Let us compute the momentum for $\tilde{t}>0$, $P=\frac{\sqrt{\lambda}}{2\pi}\mathcal{P}$
\begin{eqnarray}
\mathcal{P}(\tilde{t})&=& \int_{-\tilde{t}}^{\tilde{t}}d \tilde{x} \frac{\partial L}{\partial \dot{\tilde{x}}}=\int_{-\tilde{t}+\epsilon_0}^{\tilde{t}-\epsilon_0}d \tilde{x} \frac{\dot{\tilde{z}}\tilde{z}'}{\tilde{z}^2 \sqrt{1- \dot{\tilde{z}}^2+ \tilde{z}^{'^2}}}
\nonumber\\
&=&-\frac{\mu \tilde{t}}{2 \sqrt{2}\epsilon_0}+\frac{\mu}{4\sqrt{2}}\bigg[-1 + \frac{2}{1+ 2 \mu^2 \tilde{t}^2}- \log \frac{\epsilon_0^2}{4 \tilde{t}^2}\bigg]+O(\epsilon_0)
\end{eqnarray}
For $\tilde{t}<0$ we obtain
\begin{equation}
\mathcal{P}(\tilde{t})=\frac{\mu \tilde{t}}{2 \sqrt{2}\epsilon_0}+\frac{\mu}{4\sqrt{2}}\bigg[-1 + \frac{2}{1+ 2 \mu^2 \tilde{t}^2}- \log \frac{\epsilon_0^2}{4 \tilde{t}^2}\bigg]+O(\epsilon_0)
\end{equation}
At $\tilde{t}=\pm \infty$ the energy of the string is the same, while at intermediate times the energy increases. The same happens with the momentum of the string. Thus when the gluon and quark get closer together the energy of the system increases. While both the energy and momentum of string depend on time $\tilde{t}$, their difference does not (here we ignore the linear divergent terms)
\begin{equation}
E=P - \frac{\sqrt{\lambda}}{4 \sqrt{2}\pi}\mu
\end{equation}

 As we already pointed out, there is a time delay when the rounded spike moves from $\tilde{t}\rightarrow -\infty$ to $\tilde{t}\rightarrow \infty$ given by $\triangle \tilde{t}= 2 \delta_1= \frac{2 \sqrt{2}}{\mu}$. In $1+1$ dimensional theories this time delay translates into a phase-shift which semiclassically  is related to the time delay as \cite{jw}
\begin{equation}
\frac{d \delta(E)}{d E}=\frac{1}{2}\triangle \tilde{t}(E)  \label{pha}
\end{equation}
To compute this phase shift we need first to get the energy of the incoming gluon. We do not know how to define the energy of the incoming gluon but,
since the only scale of the problem is $\mu$ we can assume that $E\sim \mu$. This gives then a phase shift $\delta\sim \ln E$ for this scattering.

\subsubsection{$2$-parallel lines}

Let us look now for an open string solution which at the boundary ends on two time-like parallel lines extended in the $x_{+}$ direction as (we assume $b_1< b_2$)
\begin{equation}
x_{+}=\tau, \quad x_{-}=0, \quad x_1= b_1, \quad x_2=0
\end{equation}
\begin{equation}
x_{+}=\tau, \quad x_{-}=0, \quad x_1= b_2, \quad x_2=0
\end{equation}
The full ansatz for such a solution is
\begin{equation}
x_{+}=\tau, \quad x_{-}=0, \quad x_1= \sigma, \quad z=z(\sigma), \quad x_2=0
\end{equation}
We want to have the condition at the boundary $z( b_1)=z(b_2)=0$.
The Nambu-Goto Lagrangian is explicitly dependent on $\sigma$
\begin{equation}
S= \frac{\sqrt{\lambda}}{2 \pi}\int d \tau d \sigma \frac{\mu}{z^2}\sqrt{(\sigma^2+ z^2)(1+ z'^2)}  \label{act}
\end{equation}
The equation of motion for $z$ is
\begin{equation}
z'' z (z^2 + \sigma^2) + (1+ z'^2) ( z^2 + \sigma z z'+ 2 \sigma^2)=0
\end{equation}
To solve this differential equation let us first consider the change in function $z= \sigma f(\sigma)$, and then a change in variable $u=\ln \sigma$. Then the differential equation for $f$ is
\begin{equation}
(f'+ f'')f(f^2+1) +[1+ (f+f')^2][f^2 +f (f+f')+2]=0
\end{equation}
where here derivatives are with respect to the variable $u$. With the above change of variables we restricted $\sigma$ to positive values, but to account also for possible solutions extended to negative $\sigma$ we consider gluing different branches of solutions. Note that since $z\geq 0$, the physical solutions are those with $f\geq 0$. To proceed we consider the differential equation for $u(f)$ instead of $f(u)$ as above
\begin{equation}
 \ddot{u}=\frac{1}{f(f^2+1)}[f+ 2 (1+ 2 f^2) \dot{u}+ 6 f(1+f^2) \dot{u}^2+ 2 (1+ f^2)^2 \dot{u}^3]
 \end{equation}
Setting $y=\dot{u}$, we integrate the equation for $y(f)$ and obtain the solution
\begin{equation}
y(f)=-\frac{f}{1+f^2}\pm \frac{f^2}{(1+f^2)\sqrt{1+2 f^2+c (1+f^2)^2}} \label{mwl}
\end{equation}
where $c$ is an integration constant. For $c\geq 0$ the expression under the square root is always positive. For $c< 0$ we need  $c>-1$, and $f$ in the interval $0\leq f \leq \sqrt{-\frac{1+c+\sqrt{1+c}}{c}}$ in order for the square root to be real.

The expression in (\ref{mwl}) can be further integrated and we obtain a solution for $u=u(f)$ in terms of Elliptic integrals. The boundary condition for the last integration is $u(0)=\ln b_2$ with $b_2>0$. After the integration the result can be expressed as a transcendental equation for $z=z(\sigma)$ as
$\sigma= e^{u(\frac{z}{\sigma})}$. However, the other boundary condition that we want is that the other end of the string ends also on the boundary.  Therefore we want to have $z= f e^{u(f)} \rightarrow 0$ for some $f$. It turns out that this is only possible when the constant of integration $c=0$ and only for the minus sign solution in (\ref{mwl}). Below we see that we can have solutions for arbitrary $c>0$, which are extended to the negative $\sigma$ region. For $c<0$ there are no solutions possible even if one tries to glue solutions because the two branches in (\ref{mwl}) behave differently.

Let us analyze first the solution with $c=0$. In this case in the limit $f\rightarrow \infty$, indeed we have $z\rightarrow 0$, and this happens for $\sigma=b_1=0$.
\begin{figure}
\centerline{\includegraphics[scale=1]{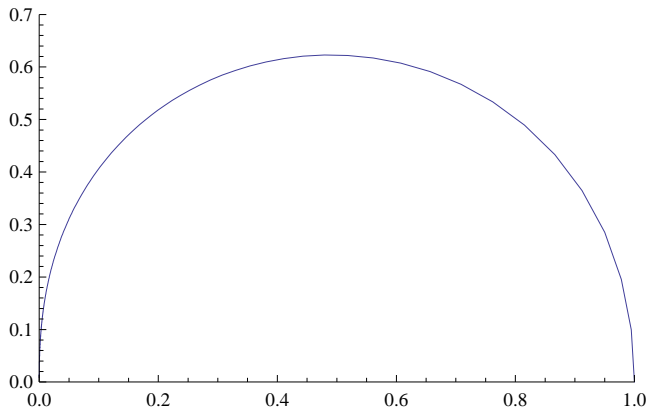}}\caption{Plot of $z$ versus $\sigma$ for $b_2=1$.}
\label{fig3}
\end{figure}
For $c=0$ the transcendental equation for $f=f(\sigma)$ simplifies to
\begin{equation}
\sigma= \frac{b_2}{\sqrt{1+f^2}}\bigg(\frac{\sqrt{1+2 f^2}-f}{\sqrt{1+2 f^2}+f}\bigg)^{-\frac{1}{2}}(\sqrt{2}f+\sqrt{1+2 f^2})^{-\frac{1}{\sqrt{2}}} \label{khj}
\end{equation}
 We plot numerically the solution $z=z(\sigma)$ in figure \ref{fig3}. This solution is independent of $\mu$. The constant $b_2$ can be expressed as
\begin{equation}
b_2= \int_0^{b_2}d \sigma = - \int_0^{\infty} d f \sigma' \label{nkk}
\end{equation}
where $\sigma'=\frac{d \sigma}{d f}$ and one needs to use (\ref{khj}) and integrate this expression. The integral cannot be done exactly, but numerically one can shown that relation (\ref{nkk}) is satisfied. Note that $b_2$ drops out from (\ref{nkk}) therefore $b_2$ is not fixed. The value of $f$ when $z$ is maximum can be obtained from
\begin{equation}
\frac{d z}{d \sigma}= -\frac{1}{f}- f + \sqrt{1+2 f^2}=0 \quad  \Rightarrow f_0= \sqrt{\frac{1+\sqrt{5}}{2}}
\end{equation}
so that the position of the maximum in $\sigma$ is
\begin{equation}
\sigma_0= b_2 \sqrt{\frac{1+\sqrt{5}}{2}}[\sqrt{1+\sqrt{5}}+\sqrt{2+\sqrt{5}}]^{-\frac{1}{\sqrt{2}}} \simeq 0.4897 \ b_2
\end{equation}
Notice that the position of the maximum $z$ is not at half of the $\sigma$ interval.

\begin{figure}
\centerline{\includegraphics[scale=1]{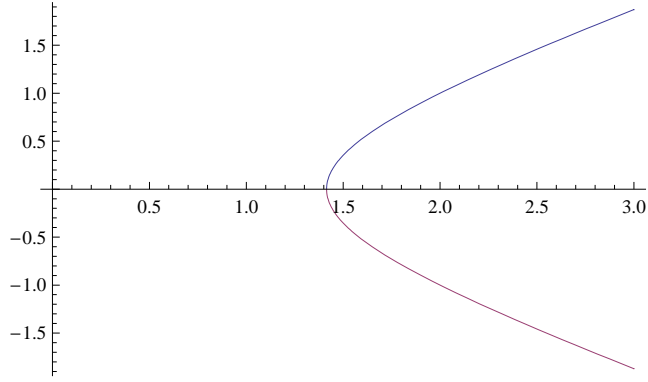}}\caption{Plot of $\tilde{t}$ versus $\tilde{x}_1$ for $b_2=\sqrt{2}$ at the boundary $z=0$.}
\label{fig4}
\end{figure}

This solution ends on the boundary at $\sigma=0$ on a light-like line and at $\sigma=b_2$ on a time-like curve. Let us see how these lines look in the pure $AdS_5$ metric. The line at $\sigma=b_2$ maps into
\begin{equation}
\tilde{x}_{+}= \frac{1}{\mu}\tan \mu \tau, \quad \tilde{x}_{-}= -\frac{1}{2} \mu b_2^2 \tan \mu \tau, \quad \tilde{x}_1=  \frac{b_2}{\cos \mu \tau}, \quad \tilde{x}_2=0  \label{lgs}
\end{equation}
while the line at $\sigma=0$ maps into
\begin{equation}
\tilde{x}_{+}= \frac{1}{\mu}\tan \mu \tau, \quad \tilde{x}_{-}= 0, \quad \tilde{x}_1=  0, \quad \tilde{x}_2=0  \label{lgs2}
\end{equation}
The last null line is rather simple. The former line may be expressed as
\begin{equation}
\tilde{x}_1= b_2 \sqrt{1+ \mu^2 \tilde{x}_{+}^2}, \quad \tilde{x}_{-} = -\frac{1}{2}b_2^2 \mu^2 \tilde{x}_{+}
\end{equation}
While in the $pp$-wave $AdS_5$, in the $(x_1,x_{+})$ plane the Wilson loop were straight parallel lines, in the pure $AdS_5$ one remains straight line but the other is a hyperbola in the $(\tilde{x}_1,\tilde{x}_{+})$ plane.  As these Wilson lines might correspond to physical particles moving on the boundary let us consider plotting $\tilde{t}$ versus $\tilde{x}_1$. Let us choose $b_2^2=2$ in (\ref{lgs}). Then after a boost in the $\tilde{x}_{+},\tilde{x}_{-}$ directions one can get rid of the factors of $\mu$ in front, so the hyperbola line is
\begin{equation}
\tilde{x}_{+}= \tan \mu \tau, \quad \tilde{x}_{-}= - \tan \mu \tau, \quad \tilde{x}_1= \frac{\sqrt{2}}{\cos \mu \tau}, \quad \tilde{x}_2=0
\end{equation}
Therefore one obtains
\begin{equation}
\tilde{t}=\sqrt{2} \tan \mu \tau, \quad \tilde{x}=0, \quad \tilde{x}_1= \sqrt{2 (1+ \tilde{t}^2)}  \label{lnf}
\end{equation}
while for the light-like line
\begin{equation}
\tilde{t}=\tilde{x}=\frac{1}{\sqrt{2}}\tan \mu \tau, \quad \tilde{x}_1=0
\end{equation}
The hyperbola $\tilde{t}=\pm \sqrt{\frac{\tilde{x}_1^2}{2}-1}$ corresponding to the time-like line (\ref{lnf}) is represented in figure \ref{fig4}.

The conserved charges in the $pp$-wave $AdS_5$ background can be expressed as
\begin{equation}
P_{+}=\frac{\sqrt{\lambda}}{2 \pi}\mu \int_{0}^{b_2} d \sigma \frac{\sqrt{(\sigma^2+z^2)(1+z'^2)}}{z^2}=-\frac{\sqrt{\lambda}}{2 \pi}\mu \int_0^{\infty}d f \frac{\sigma'}{\sigma f^2}\sqrt{(1+f^2)(1+ (f+ \frac{\sigma}{\sigma'})^2)}\label{pp}
\end{equation}
\begin{equation}
P_{-}=-\frac{\sqrt{\lambda}}{2 \pi \mu}\int_{0}^{b_2} d \sigma \frac{1}{z^2}\sqrt{\frac{1+z{'^2}}{\sigma^2+z^2}}=\frac{\sqrt{\lambda}}{2 \pi \mu}\int_0^{\infty}d f \frac{\sigma'}{\sigma^3 f^2}\sqrt{\frac{1+(f+\frac{\sigma}{\sigma'})^2}{1+f^2}}\label{pm}
\end{equation}
 Near the boundary of $AdS_5$ the integrals diverge, therefore we introduce a cutoff $\epsilon$ in $z$. While we cannot perform the above integrals exactly we can extract the divergent parts. Cutoff in $z$ translates into cutoffs in $f$. Recall that $z=0$ both for $f=0$ and for $f\rightarrow\infty $, thus there is a $\epsilon$ dependent large small cutoff in $f$, $f_{min}$ as well as a large one $f_{max}$. Using (\ref{khj}) and let us expand $z= \sigma f$ in the limits
\begin{equation}
z=\sigma f = b_2 f+ O(f^2) = \epsilon  \quad \Rightarrow f_{min}=\frac{\epsilon}{b_2}
\end{equation}
\begin{equation}
z=\sigma f=f^{-\frac{1}{\sqrt{2}}}[ b_2 (1+ \sqrt{2})2^{-\frac{3}{2 \sqrt{2}}} +O(\frac{1}{f^2})]=\epsilon \quad \Rightarrow f_{max}=\bigg(\frac{b_2}{\epsilon}\bigg)^{\sqrt{2}}\frac{(1+\sqrt{2})^{\sqrt{2}}}{2\sqrt{2}}
\end{equation}
To compute the charges (\ref{pp},\ref{pm}) we expand the integrands in the small and large $f$ limits and extract only the values of the integral near $f_{min}$ and $f_{max}$. More precisely (\ref{pp}) becomes
\begin{equation}
P_{+}=-\frac{\sqrt{\lambda}\mu}{2 \pi}\bigg[\int_{f_{min}}d f (-\frac{1}{f^2})+ \int^{f_{max}}(-\frac{1}{\sqrt{2}f})\bigg]=\frac{\sqrt{\lambda}\mu}{2 \pi}\bigg(\frac{b_2}{\epsilon}+\log \frac{b_2}{\epsilon}+O(\epsilon^0)\bigg)
\end{equation}
For the other charge (\ref{pm}) we obtain
\begin{equation}
P_{-}=\frac{\sqrt{\lambda}}{2 \pi \mu}\bigg[\int_{f_{min}}d f (-\frac{1}{b_2^2 f^2})+\int^{f_{max}}d f \frac{(2 \sqrt{2}-3)2^{\frac{3}{\sqrt{2}}-\frac{1}{2}}}{b_2^2}f^{\sqrt{2}-1}\bigg]=-\frac{\sqrt{\lambda}}{2 \pi \mu}\bigg[\frac{1}{2\epsilon^2}+\frac{1}{b_2 \epsilon}+O(\epsilon^0)\bigg]
\end{equation}
Keeping the leading orders in $\epsilon$ in these charges, we eliminate $\epsilon$ and obtain the relationship between the divergent parts of the charges
\begin{equation}
P_{-}\approx - \frac{\pi}{\sqrt{\lambda}\mu^3 b_2^2} P_{+}^2
\end{equation}
Let us finish the discussion about the charges by pointing out that the action (\ref{act}) on the solution is real, and up to the length of the $\tau$ interval it is the same as $P_{+}$.

\bigskip

\bigskip

\begin{figure}
\centerline{\includegraphics[scale=1]{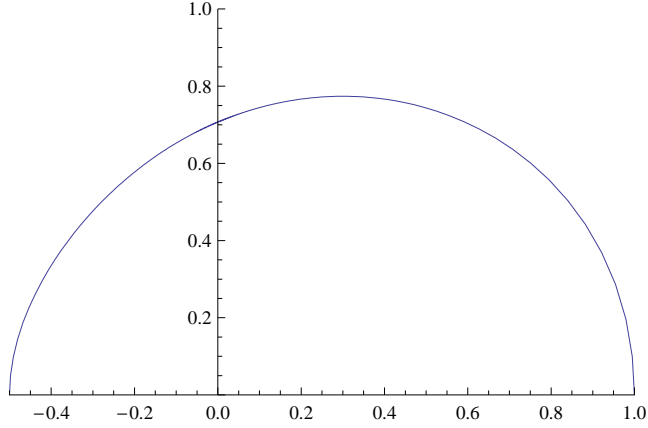}}\caption{Plot of $z$ versus $\sigma$ for $b_2=1$, $b_1=0.5$, $c=4.77$.}
\label{fig5}
\end{figure}

Let us return to the differential equation (\ref{mwl}) and consider the situation $c>0$. The two branches in eq.(\ref{mwl}) behave in the same way. They both satisfy $\sigma \rightarrow 0$ as $f\rightarrow \infty$, but there is no point other then $\sigma=b$ for which $z=0$. We take the negative branch, denoted $z_{{-}}$, to depend on $c$ and $b_2>0$, while the positive one, $z_{{+}}$ on $c$ and $b_1>0$. For fixed $b_1$ and $b_2$ we want to find a solution for $c=c(b_1,b_2)$ so that
\begin{equation}
z_{+}-z_{-}\bigg|_{f\rightarrow\infty}=0, \quad \frac{d z_{+}}{d \sigma}+\frac{d z_{-}}{d \sigma}\bigg|_{f\rightarrow \infty}=0  \label{amf}
\end{equation}
For such solutions we use the positive branch for $\sigma<0$, and therefore this branch smoothly continues to the negative branch to negative values of $\sigma$ until it reaches the boundary at $\sigma=-b_1$. Let us find the constant $c$ satisfying (\ref{amf}). The solutions of (\ref{mwl}) for $c>0$ is (recall that $\sigma=e^{u}$)
\begin{eqnarray}
u_{\pm}(f)&=&\mp G(f,c)+\ln \frac{b}{\sqrt{1+f^2}}\label{soll}
\end{eqnarray}
where
\begin{eqnarray}
G(f,c)&=&\frac{i}{\sqrt{1+c-\sqrt{1+c}}}\bigg(F[i \sinh^{-1} [f \sqrt{1-\frac{1}{\sqrt{1+c}}}],\frac{2+c+2 \sqrt{1+c}}{c}]\nonumber\\
&-&\Pi[\frac{1+c +\sqrt{1+c}}{c},i \sinh^{-1} [f \sqrt{1-\frac{1}{\sqrt{1+c}}}],\frac{2+c +2 \sqrt{1+c}}{c}]\bigg)
\end{eqnarray}
where $F[x,m]$ and $\Pi[n,x,m]$ are the incomplete Elliptic functions defined as
\begin{equation}
F[x,m]=\int_{0}^{x}\frac{d \theta}{\sqrt{1- m \sin^2 \theta}}, \quad \Pi[n,x,m]=\int_{0}^{x}\frac{d \theta}{(1- n \sin^2 \theta) \sqrt{1-m \sin^2 \theta}}
\end{equation}
The conditions (\ref{amf}) become
\begin{eqnarray}
&&\frac{1}{2}\ln \frac{b_1}{b_2}=G(f,c)\bigg|_{f\rightarrow \infty}\label{wfs}\\
&=&\sqrt{\frac{1+c+\sqrt{1+c}}{c}}\bigg[-K[-\frac{2 (1+\sqrt{1+c})}{c}]+\Pi[\frac{1}{1-\sqrt{1+c}},-\frac{2(1+\sqrt{1+c})}{c}]\bigg]\nonumber
\end{eqnarray}
and
\begin{equation}
2f + \frac{1}{u'_{+}(f)}+\frac{1}{u'_{-}(f)}\bigg|_{f\rightarrow\infty}=- \frac{1+c}{c}\frac{2}{f}+O(\frac{1}{f^2})=0 \label{ngm}
\end{equation}
where $K[x]=F[\frac{\pi}{2},x]$ and $\Pi[x,y]=\Pi[x,\frac{\pi}{2},y]$ are Elliptic integrals. The condition (\ref{ngm}) is satisfied in the limit $f\rightarrow\infty$. The right hand side in (\ref{wfs}) is negative and satisfies $\lim_{c \rightarrow 0}G(\infty,c)=-\infty$, and $\lim_{c\rightarrow \infty}G(\infty,c)=0.$
The fact that $G(\infty,c)$ is negative is in accord with our initial assumption $b_1<b_2$. Also, the limit $b_1\rightarrow 0$ gives $c \rightarrow 0$ in accord to the previous $c=0$ analysis. For a fixed ratio $\frac{b_1}{b_2}<1 $, equation (\ref{wfs}) gives a solution for $c=c(b_1,b_2)$. While it is not possible to invert (\ref{wfs}) to get $c=c(b_1,b_2)$ one can solve it numerically. For example for $b_2=1$, $b_1=0.5$ we obtain $c\simeq 4.77$. Gluing together the two solutions as explained above we obtain the solution shown in figure \ref{fig5}.
On the boundary ($z=0$) this solution ends on two time-like lines at $x_1=b_2$ and $x_1=-b_1$. They can be mapped to pure $AdS_5$ coordinates as in (\ref{lgs}) where the lines look like hyperbolas.

Let us compute the charges (\ref{pp},\ref{pm}) in this case with $c>0$. Again the integrals cannot be computed exactly but we can extract the divergent terms coming from the region near the boundary $z=0$. We need to expand the two branches given in (\ref{soll}) near $f=0$ using the negative branch near $\sigma=b_2$ and the positive branch near $\sigma=b_1$. This leads to two cutoffs in $f$ near the boundary of $AdS_5$
\begin{equation}
z_{+}= f \sigma = b_1 f+ O(f^2)=\epsilon \quad \Rightarrow f_1=\frac{\epsilon}{b_1},
\end{equation}
and
\begin{equation}
z_{-}= f \sigma = b_2 f+ O(f^2)=\epsilon \quad \Rightarrow f_2=\frac{\epsilon}{b_2}
\end{equation}
The charge (\ref{pp}) can be written as
\begin{eqnarray}
P_{+}&=&-\frac{\sqrt{\lambda}}{2 \pi}\mu \bigg[\int_{f_{1}}^{\infty}d f \frac{\sigma_{+}'}{\sigma_{+} f^2}\sqrt{(1+f^2)(1+ (f+ \frac{\sigma_{+}}{\sigma_{+}'})^2)}\nonumber\\
&+&\int_{f_{2}}^{\infty}d f \frac{\sigma_{-}'}{\sigma_{-} f^2}\sqrt{(1+f^2)(1+ (f+ \frac{\sigma_{-}}{\sigma_{-}'})^2)}\bigg]
\end{eqnarray}
Expanding the integrands near $f=0$ we obtain
\begin{equation}
P_{+}=-\frac{\sqrt{\lambda}}{2 \pi}\mu \bigg[\int_{f_{1}}d f (-\frac{1}{f^2})+\int_{f_{2}}d f (-\frac{1}{f^2})\bigg]=\frac{\sqrt{\lambda}}{2 \pi}\mu \frac{b_1+b_2}{\epsilon}+O(\epsilon^0)
\end{equation}
The other conserved charge (\ref{pm}) can be computed in a similar way and we obtain
\begin{equation}
P_{-}=-\frac{\sqrt{\lambda}}{2 \pi \mu}\frac{b_1+b_2}{b_1 b_2 \epsilon}+O(\epsilon^{0})
\end{equation}
Let us observe that the divergent part of the charges do not depend explicitly on $c$. Eliminating $\epsilon$ we obtain the following relationship between the divergent parts of the charges
\begin{equation}
\frac{P_{+}}{P_{-}} \approx -\mu^2 b_1 b_2
\end{equation}

\subsection{Straight lines along the time direction $t$}

\subsubsection{$1$-line}

Let us start with the metric (\ref{owd}) restoring the time coordinate in Poincare patch ($t=x_0$)
\begin{equation}
ds^2= \frac{1}{z^2}(- d t^2 + d x^2 - \frac{\mu^2}{2}(x_i^2+ z^2) (d x + dt)^2 + dx_i^2 + dz^2)
\end{equation}
We look for a solution which at the boundary ends on a time-like line along the $t$ direction. A solution of this form is
\begin{equation}
t=\tau, \quad x=0, \quad z=\sigma, \quad x_i=0
\end{equation}
Here $\sigma>0$.
In the $pp$-wave $AdS_5$ space we compute the conserved charges and obtain
\begin{equation}
E= \frac{\sqrt{\lambda}}{2 \pi}\bigg[\frac{1}{\epsilon}+ \frac{\mu}{\sqrt{2}}\log R + O(R^{0},\epsilon)\bigg]
\end{equation}
where we have introduced a cutoff $\epsilon$ near the boundary and a cutoff $R$ near the horizon. Here cutoffs in $z$ are the same as cutoffs in $\sigma$. The other conserved charge corresponding to the isometry along $x$ is
\begin{equation}
P= \frac{\sqrt{\lambda}}{2 \pi}\bigg[\frac{\mu}{\sqrt{2}}\log R+ O(R^{0})\bigg]
\end{equation}

Computing the action on this solution we obtain the same expression as the energy up to a factor of $T$, where we take $-\frac{T}{2}\leq \tau \leq\frac{T}{2}$ with $T$ large
\begin{equation}
S= \frac{\sqrt{\lambda}}{2 \pi}T \int_{\epsilon}^{R}d \sigma \frac{1}{\sigma^2}\sqrt{1+ \frac{\mu^2}{2}\sigma^2}= \frac{\sqrt{\lambda}}{2 \pi}T\bigg[\frac{1}{\epsilon}+ \frac{\mu}{\sqrt{2}}\log R + O(R^{0},\epsilon)\bigg] \label{mah}
\end{equation}
We observe that the only divergency in $\epsilon$ is linear; like in the previous cases this divergency should be canceled by a boundary term. In the limit $\mu \rightarrow 0$ the $pp$-wave space is the same as pure $AdS_5$, and we see that in this limit (\ref{mah}) reduces to the well known straight string solution \cite{dgt}.

\begin{figure}
\centerline{\includegraphics[scale=1.0]{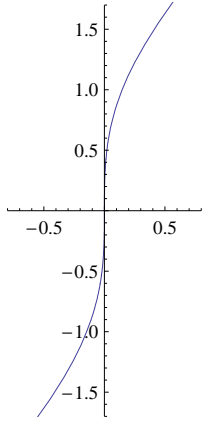}}\caption{Parametric plot of $\tilde{t}$ versus $\tilde{x}$, for $\mu=\sqrt{2}$ at the boundary $z=0$.}
\label{fig6}
\end{figure}

\begin{figure}
\centerline{\includegraphics[scale=1.0]{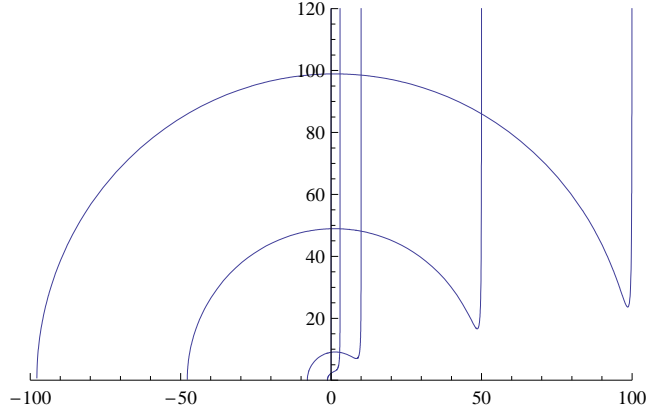}}\caption{Parametric plot of $\tilde{z}$ versus $\tilde{x}$, for $\mu=1$, and $\tilde{t}=-100,-50,-10,-3$.}
\label{fig7}
\end{figure}

Let us now map this solution to pure $AdS_5$. Under the map (\ref{tra}) this solution is mapped into
\begin{equation}
\tilde{x}_{+}=\frac{1}{\mu}\tan \frac{\mu \tau}{\sqrt{2}}, \quad \tilde{x}_{-}= - \frac{\tau}{\sqrt{2}}-\frac{\mu}{2}\sigma^2 \tan \frac{\mu \tau}{\sqrt{2}}, \quad \tilde{z}=\frac{\sigma}{\cos \frac{\mu \tau}{\sqrt{2}}}, \quad \tilde{x}_i=0
\end{equation}
On the boundary the straight line in $pp$-wave coordinates is mapped into a line with tangent shape in pure $AdS_5$ as shown in Figure \ref{fig6}.
The world-sheet surface is described by the equation
\begin{equation}
\tilde{z}=\sqrt{- \frac{2}{\mu^2 \tilde{x}_{+}}(\tilde{x}_{-}+\frac{\arctan \mu \tilde{x}_{+}}{\mu})(1+\mu^2 \tilde{x}_{+}^2)} \label{adq}
\end{equation}
This looks similar to the surface in (\ref{lnd}) as one can see in figure \ref{fig7} for $\tilde{t}<0$.
The difference with the case in (\ref{lnd}) is that now the point where the line ends at the boundary is shifted compared to the situation in (\ref{lnd}). It is obtained from the solution of equation $\tilde{x}_{-}+\frac{\arctan \mu \tilde{x}_{+}}{\mu}=0$. This shows that in the present case the end point on the boundary moves with a speed less than the speed of light. The position of the vertical line is obtained when $\tilde{x}_{+}=0$, i.e. $\tilde{x}=-\tilde{t}$. On the worldsheet, the process can again be viewed as an elastic scattering process, while in the spacetime/gauge theory it can be viewed as the scattering between a gluon and a quark moving at a speed less than the speed of light.

Following what was done for the line in the $x_{+}$ direction in (\ref{jes}), we would like to compute the energy and momentum of string. However, in this case the computation cannot be done exactly because of the complicated form of the solution  (\ref{adq}), and the fact that the equation $\tilde{x}_{-}+\frac{\arctan \mu \tilde{x}_{+}}{\mu}=0$ cannot be solved explicitly for $\tilde{x}_0=\tilde{x}_0(\tilde{t})$. This solution gives the limit of integration in the computation of energy which for $\tilde{t}>0$ is $-\tilde{t}\leq \tilde{x}\leq \tilde{x}_0$. As in the case of the line in the $x_{+}$ direction, here we again expect that both energy and momentum of the string depend on time. The time delay for the rounded spike as it moves from one site (at $\tilde{t}=-\infty$ ) to the other (at $\tilde{t}=\infty$) of the vertical line can be computed and we obtain $\triangle \tilde{t}=2 \delta_1= \frac{2 \sqrt{2}}{\mu}$. This is the same as what we had in the case of the solution in the $x_{+}$ direction in section 3.1.1. Since $\mu$ is the only physical scale available we expect that asymptotically at $|\tilde{t}|\rightarrow \infty$, the string energy will be again proportional to $\mu$. Upon eliminating $\mu$ we expect that in this case the resulting phase shift is again of the form $\delta(E) \sim \ln E$.

 It is also interesting to consider a solution obtained from a particle which, in the pp-wave, moves at constant velocity. It interpolates between
the one at rest, considered here and the one moving at the speed of light considered in the previous subsection.
The solution in flat \ads{5} space can be obtained simply by using the symmetry:
\beq
 \tilde{x}_+ \rightarrow \tilde{x}_+, \ \ \tilde{x}_-\rightarrow \frac{1}{\xi^2} \tilde{x}_-, \ \ \tilde{x}_i \rightarrow \frac{1}{\xi}\tilde{x}_i,\ \
\tilde{z} \rightarrow \frac{1}{\xi} \tilde{z}
\eeq
which gives, from (\ref{adq}), the new solution:
\beq
\tilde{z}=\sqrt{- \frac{2}{\mu^2 \tilde{x}_{+}}(\tilde{x}_{-}+\xi^2 \frac{\arctan \mu \tilde{x}_{+}}{\mu})(1+\mu^2 \tilde{x}_{+}^2)} \label{adq2}
\eeq
interpolating between eq.(\ref{adq}) for $\xi=1$ and eq.(\ref{lnd}) for $\xi=0$.  The boundary Wilson loop is described by the equation
\beq
 \tilde{x}_{-}= - \xi^2 \frac{\arctan \mu \tilde{x}_{+}}{\mu}
\eeq
 It represents a particle coming from infinity at the speed of light which slows down, attaining its minimum speed
\beq
 v =\left. \frac{1+\partial_+ \tilde{x}_-}{1-\partial_+ \tilde{x}_-} \right|_{\tilde{x}_+=0} = \frac{1-\xi^2}{1+\xi^2}
\eeq
at $\tilde{x}=\tilde{t}=0$. We see that the particle is at rest at the origin when $\xi=1$ and moves at the speed of light when $\xi=0$.

\subsubsection{$2$-parallel lines}

We extend the above construction and look for a solution which at the boundary $z=0$ consists of two time-like parallel lines along the $t$ direction (let us assume $b > 0$)
\begin{equation}
t=\tau, \quad x= \pm b, \quad x_i=0
\end{equation}
The ansatz for such a solution is
\begin{equation}
t =\tau, \quad x=\sigma, \quad z=z(\sigma), \quad x_i=0  \label{sol}
\end{equation}
The Nambu action becomes
\begin{equation}
S=\frac{\sqrt{\lambda}}{2 \pi}\int d \tau d \sigma \frac{1}{z^2}\sqrt{1+ (1+ \frac{\mu^2}{2}z^2)z'^2}
\end{equation}
We want the solution for $z$ to obey the boundary condition $z(\pm b)=0$. Since the action does not depend explicitly on $\sigma$ the conserved charge is
\begin{equation}
\frac{1}{z^2}\frac{1}{\sqrt{1+ (1+ \frac{\mu^2}{2}z^2)z'^2}}= d^2
\end{equation}
where $d$ is a constant. The differential equation obeyed by $z$ is
\begin{equation}
z' = \pm \frac{1}{z^2 d^2}\sqrt{\frac{1- d^4 z^4}{1+ \frac{\mu^2}{2}z^2}} \label{leg}
\end{equation}
The worldsheet surface reaches a maximum value in the radial direction of $AdS_5$, i.e. $0 \leq z \leq \frac{1}{d}$, where we assume $d > 0$. The distance between the Wilson lines at the boundary can be expressed as
\begin{equation}
2 b = \int_{-b}^{b}d \sigma= 2 d^2 \int_0^{1/d} d z \ z^2 \sqrt{\frac{1+ \frac{\mu^2}{2}z^2}{1- d^4 z^4}} \label{jfb}
\end{equation}
Let us note that in the limit $\mu \rightarrow 0$ the equation of motion reduces, as it should, to the equation corresponding to the usual two parallel lines \cite{dgt} solution
\begin{equation}
z' = \pm \frac{1}{z^2 d^2}\sqrt{1- d^4 z^4} \label{leg3}
\end{equation}

\begin{figure}
\centerline{\includegraphics[scale=0.5]{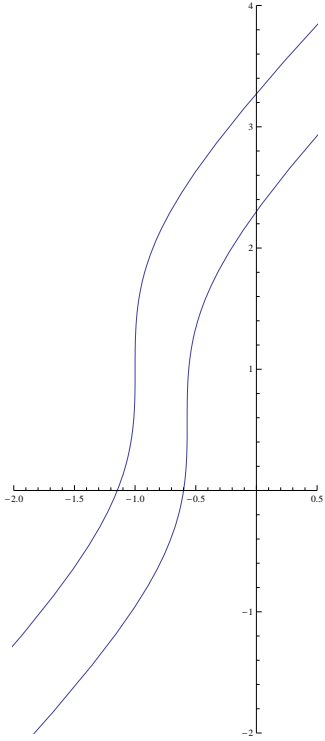}}\caption{Parametric plot of $\tilde{t}$ versus $\tilde{x}$, for $\mu=\sqrt{2}$ and  $b=1$ at the boundary $z=0$. The upper curve is the image in the tilde coordinates of the $x=- b$ straight line, and the lower curve corresponds to the $x=b$ straight line.}
\label{fig8}
\end{figure}

It is possible to express the integrals of the equation of motion (\ref{leg}) and condition (\ref{jfb}) in general for arbitrary values of $\mu, d$, in terms of the elliptic integrals but the expressions are not very useful, so we do not write them down. Only two of the parameters $b,\mu,d$ are independent because of the condition (\ref{jfb}). Let us see the form of the solution after considering the transformation (\ref{tra}) that brings the metric back into pure $AdS_5$. The explicit solution is complicated in the general case, but we can look at how the Wilson lines at the boundary ($z=0$) transform
\begin{equation}
\tilde{x}_{+}= \frac{1}{\mu}\tan \frac{\mu}{\sqrt{2}}( \pm b + \tau), \quad \tilde{x}_{-}=x_{-}=\frac{1}{\sqrt{2}}(\pm b-\tau), \quad \tilde{x}_i=0
\end{equation}
In the pure $AdS_5$ coordinates, $\tilde{t}, \tilde{x}$ become
\begin{equation}
\tilde{t}=\frac{1}{\sqrt{2}\mu}\tan \frac{\mu}{\sqrt{2}}(\pm b +\tau)-\frac{1}{2}(\pm b -\tau), \quad \tilde{x}=\frac{1}{\sqrt{2}\mu}\tan \frac{\mu}{\sqrt{2}}(\pm b +\tau)+\frac{1}{2}(\pm b -\tau)
\end{equation}
While in the $pp$-wave $AdS_5$ space, in the plane $t,x$ we had two straight lines in the $t$ direction, in the pure $AdS_5$ in the plane $\tilde{t},\tilde{x}$ we have more complicated curves. The curves at $x= \pm b$ are plotted in the new coordinates in figure \ref{fig8}. As expected, of course, the curves in the pure $AdS_5$ coordinates are again  time-like.

Let us compute the conserved charges associated to translations in $t$ and $x$ in the $pp$-wave $AdS_5$ background. The energy of the string is
\begin{equation}
E= \frac{\sqrt{\lambda}}{2 \pi}\int d \sigma \frac{\partial L}{\partial(\partial_{\tau}t)}= \frac{\sqrt{\lambda}}{2 \pi} 2 \int_{\epsilon}^{{\frac{1}{d}}} dz \frac{1}{z^2} \sqrt{\frac{1+\frac{\mu^2}{2}z^2}{1- d^4 z^4}} \label{lml}
\end{equation}
where we introduced a cutoff at small $z$. Let us recall that in fact $\mu$ can be set to $1$ by rescaling of $x_{+}$ and $x_{-}$. For simplicity let us set $\mu = 1$. The integral in (\ref{lml}) can be done and, expanding the result in small $\epsilon$ we obtain
\begin{equation}
E=\frac{\sqrt{\lambda}}{ \pi} \frac{1}{\epsilon}+\frac{\sqrt{\lambda}}{2 \sqrt{2} \pi} \frac{1}{d}\bigg[(1+ 2 d^2) K[\frac{2 d^2-1}{4 d^2}]-4 d^2 E[\frac{2 d^2-1}{4 d^2}]\bigg]+O(\epsilon)
\end{equation}
where $E[x]$, $K[x]$ are elliptic integrals.
The other conserved charge is
\begin{equation}
P= \frac{\sqrt{\lambda}}{2 \pi}  \int d \sigma \frac{\partial L}{\partial(\partial_{\tau}x)}=\frac{\sqrt{\lambda}}{2 \pi} \frac{\mu^2}{2} 2 \int_{0}^{{\frac{1}{d}}} dz  \sqrt{\frac{1- d^4 z^4}{1+\frac{\mu^2}{2}z^2}}
\end{equation}
This integral can be expressed in terms of Elliptic integrals but the exact expression is not very illuminating. Let us mention that the integral is convergent so here we do not need a cutoff in $z$. The charge $P$ should be dependent on the length of the string in the $x$ direction, thus providing a relation between the parameters $\mu,d,b$. Of course, we already exploited such a relation in (\ref{jfb}).

We observe that for a particular value of $d$, i.e. $d=\frac{1}{\sqrt{2}}$ (or $d=\frac{\mu}{\sqrt{2}}$ for arbitrary $\mu$)  the integration in (\ref{leg}) is simpler and the solution, in this case, can be written as a transcendental equation for $z$
\begin{equation}
- \frac{z}{\sqrt{2}} \sqrt{1- \frac{z^2}{2}}+ \arcsin \frac{z}{\sqrt{2}} +c_1 =  \pm \sqrt{2} \sigma d  \label{jen}
\end{equation}
Constant $c_1$ can be obtained from the boundary conditions that $z(\pm b)=0$. This gives $c_1= - \sqrt{2} b $. The negative branch is to be used
for positive values of $\sigma$. The condition (\ref{jfb}) can also be easily integrated and it gives an expression for $b$ as  $ b = \frac{\pi \sqrt{2}}{4}$ (or $ b = \frac{\pi \sqrt{2}}{4 \mu}$ for arbitrary $\mu$).
For this particular case the conserved charges reduce to
\begin{equation}
E= \frac{\sqrt{\lambda}}{\pi}\frac{\sqrt{1- d^2 \epsilon^2}}{\epsilon}=\frac{\sqrt{\lambda}}{ \pi} \frac{1}{\epsilon}+ O(\epsilon), \quad \quad P=\sqrt{\lambda}\frac{d}{4}
\end{equation}

The computation of the action on this solution gives
\begin{equation}
S= \frac{\sqrt{\lambda}}{2 \pi} 2  T  \int_{\epsilon}^{\frac{1}{d}} d z \frac{1}{z^2}\sqrt{\frac{1+\frac{\mu^2}{2}z^2}{1- d^4 z^4}}  \label{nqu}
\end{equation}
where we have chosen the $\tau$ interval  $-\frac{T}{2}\leq \tau \leq \frac{T}{2}$, with $T$ large. Also, we introduced a small cutoff in $z$ to account for the expected divergency near the boundary of $AdS_5$. Let us recall that because of the condition (\ref{jfb}) only two parameters are independent, so the action can be expressed in general as $S=S(b,\mu)$. For simplicity let us choose again $\mu=1$. Doing the integral and expanding in small $\epsilon$ we obtain the action
\begin{equation}
S= \frac{\sqrt{\lambda}}{ \pi} \frac{T}{\epsilon}+\frac{\sqrt{\lambda}}{2 \sqrt{2} \pi} \frac{T}{d}\bigg[(1+ 2 d^2) K[\frac{2 d^2-1}{4 d^2}]-4 d^2 E[\frac{2 d^2-1}{4 d^2}]\bigg]+O(\epsilon) \label{ken}
\end{equation}
The function of $d$ at order $O(\epsilon^0)$ is decreasing. It is interesting to note that for the particular value $d=\frac{1}{\sqrt{2}}$ the O($\epsilon^0$) term vanishes. As in \cite{dgo} one needs to regularize the area by adding a boundary term that cancels the linear divergent term $\frac{1}{\epsilon}$.

So far we discussed the solution in Minkowski space. It turns out that this solution can also be found in the Euclidean space-time and worldsheet. The same was found in \cite{tz} in the case of other simple solutions that are nontrivial only in the $AdS_5$ part of $AdS_5 \times S^5$. One can go from Minkowski to Euclidean by taking $t\rightarrow i t$, $\tau \rightarrow i \tau$. The solution (\ref{sol}) is also solution in Euclidean space. The equation of motion is again (\ref{leg}). In Euclidean space it makes sense to relate the action of a string solution that ends on the boundary on a Wilson loop to the expectation value of the Wilson loop.

\section{Conclusions}
\label{conclusions}

 In this paper we have analyzed new methods to find solutions describing open strings moving in \ads{5} and ending in the boundary. According to the
AdS/CFT correspondence they are dual to Wilson loops in \N{4} SYM theory. The first solutions were found by using an ansatz leading to linear
(and unconstrained) equations of motion. They describe $\frac{1}{4}$ BPS Euclidean Wilson loops with the shape $x_+=x_+(x_1)$, $x_-=x_2=0$ in usual
Minkowski light-cone coordinates $x_\pm$, $x_{1,2}$. Using an inversion we mapped them to closed Wilson loops whose expectation value is
$\langle W\rangle= e^{-\sqrt{\lambda}}$. Other, different, solutions were found by considering Wilson loops for \N{4} SYM living in a four dimensional
pp-wave. By doing a conformal transformation we can obtain Wilson loops in usual flat space. In this way new interesting solutions were found.
They seem to describe gluons interacting with quarks, the gluons being described by spikes coming out from the horizon.
 We expect these solutions to give a basis for further generalizations and a deeper understanding of Wilson loops in superconformal theories.


\begin{acknowledgments}
We are  grateful to P. Argyres, S. Das, R. Leigh,  C. Sommerfield and C. Thorn for useful discussions. We are grateful to A. Tseytlin for
discussions and comments on the paper. M.K. and A.T. were supported in part by NSF under grant PHY-0653357.
The work of R.I. was supported in part by the Purdue Research Foundation.
\end{acknowledgments}


\end{document}